\documentclass[11pt]{article}%
\usepackage[margin=1in]{geometry}
\usepackage[english]{babel}
\usepackage[utf8]{inputenc}
\usepackage[T1]{fontenc}
\usepackage{amsmath,amssymb}
\usepackage{graphicx}
\usepackage{caption}
\usepackage{subcaption}
\usepackage{float}
\usepackage{hyperref}
\usepackage[capitalise,nameinlink]{cleveref}
\usepackage[numbers]{natbib}
\usepackage{setspace}
\usepackage{authblk}
\usepackage{xcolor}
\usepackage{ulem} 
\usepackage{todonotes}
\usepackage{comment}
\usepackage{makecell}
\usepackage{algorithm2e}

\usepackage{lineno}

\usepackage{dsfont}

\hypersetup{
colorlinks=true,
linkcolor=blue,
filecolor=blue,
urlcolor=blue,
}

\newcommand{\deleted}[1]{{\color{red}\sout{#1}}}

\title{Collaboration, Integration, and Thematic Exploration \\ in European Framework Programmes: \\ A Longitudinal Network Analysis}
\date{}

\author[1, 2]{Veronica Orsanigo}
\author[1, 4]{Thomas Louf}
\author[1]{Eleonora Andreotti}
\author[1]{Elisa Leonardelli}
\author[3]{Pierluigi Sacco}
\author[1]{Riccardo Gallotti}
\affil[1]{Fondazione Bruno Kessler, Via Sommarive 18, 38123 Povo,
Italy}
\affil[2]{University of Trento, Italy}
\affil[3]{University of Chieti-Pescara, Chieti, Italy}
\affil[4]{Universidad Carlos III de Madrid, Departamento de
Matem\'aticas, Grupo Interdisciplinar de Sistemas Complejos,
Legan\'es, Spain}

\begin{document}

\maketitle

\begin{abstract}

\noindent
Since their inception in 1984, the European Framework Programmes (FPs)
have funded collaborative R\&D to promote excellence, cohesion, and competitiveness in a growing European Union. However, their integrative impact and the evolution of the research landscape alongside its collaborative structures remain insufficiently understood. 
\noindent
In this longitudinal study, we leverage CORDIS data from all nine FPs to reconstruct the evolution of country-level collaboration networks over time. 
We observe an increasing equity in project participation between FP1 and FP6, although newly included countries systematically tend to be marginal when first joining the programmes. However, we find that the collaborative nature of EU projects progressively integrates marginal countries in the network, even if this integration is still in progress.
\noindent
We also trace the evolution in time of research topics using semantic embeddings of project descriptions, identifying 117 topics grouped into 16 macro-topics. By computing the minimum spanning tree length of project embeddings within yearly time windows, we quantify how European research progressively explores a wider knowledge space. A comparison with a null model with points randomly distributed in the semantic space indicates that this exploration is more focused than a uniform coverage. Moreover, it appears uneven, with few topics mostly attracting industry and others academia.
\noindent
Our findings suggest that, while European funding promotes international cooperation, it has not yet fully resolved core–periphery asymmetries, and European research remains concentrated along established trajectories rather than broadly exploratory, with implications for future programme design and the excellence–cohesion debate.
\end{abstract}

\textit{\textbf{Keywords -} Research collaboration network, European Framework Programmes, Network analysis, Research integration, Topic modeling, Semantic space exploration}

\section{Introduction}

The European Framework Programmes (FPs) for Research and Technological
Development represent one of the most sustained and large-scale policy
experiments in transnational scientific collaboration. Since the launch
of FP1 in 1984, nine successive programmes have channeled increasing
volumes of funding, reaching over EUR 95 billion in Horizon 2020 alone,
into collaborative R\&D projects spanning the full breadth of the
European research landscape~\cite{ec_funding}. These programmes have
served multiple, sometimes competing, policy objectives: advancing
scientific excellence on the frontier, promoting economic
competitiveness through innovation, and fostering integration and
cohesion across an expanding set of member
states~\cite{luukkonen2000,muldur2006}.

The tension between these objectives has generated a rich scholarly
debate. On the one hand, previous studies documented the dependency of ERC-funded research on previous US 
collaborations \cite{chowdhary2023dependency} and the emergence of a relatively stable core-periphery
structure in the collaboration networks in earlier FPs, with a small number of Western European countries occupying
central positions in the network~\cite{roediger2008,rodriguez2024analysing}. On
the other hand, later enlargements of the EU and deliberate policy
instruments such as ``widening'' measures in Horizon 2020 and
Horizon Europe have aimed to reduce structural asymmetries and promote
the participation of less research-intensive
countries~\cite{ec_widening}. 
At the organizational level, previous research showed that companies receiving FP funding are more active in forming regional ties than non-FP firms. However, they tend to limit collaboration among themselves, leading to weaker knowledge diffusion among firms outside international collaborations~\cite{yankova2026role}.
Whether these instruments have been
effective, and whether the progressive expansion of the network has
genuinely altered the underlying inequality of participation, remains an
open empirical question.

A parallel body of work has examined the thematic evolution of European
research, including how academic researchers switch between topics over
time~\cite{de2016quantifying}, how funding priorities shift across
programme generations~\cite{reillon2017}, and whether targeted research funding is associated with changes in the research landscape \cite{traag2026targeted}. However, this literature has
largely proceeded independently of the network-analytic strand. The
interplay between the structural dynamics of collaboration (who works
with whom, and how integration evolves) and the semantic dynamics of
research (what is being investigated, and how the thematic landscape
expands or contracts) is still relatively unexplored.

This gap matters for policy. If the expansion of the collaboration
network coincides with thematic diversification, it would suggest that
FPs are effective instruments for broadening the European research
portfolio. If, instead, new entrants are absorbed into pre-existing thematic structures without altering the overall direction of research, this would raise questions about path dependence and the capacity of FPs to promote genuinely transformational exploration.
Similarly, understanding how funding flows between research
organizations and industry across different thematic domains is critical to assessing whether FPs support balanced innovation ecosystems or reinforce sectoral asymmetries.

In the present paper, we address these questions through a comprehensive longitudinal analysis that bridges the structural and semantic dimensions of European research collaboration.

In our longitudinal study, we leverage the publicly available data about
EU projects from the CORDIS project database~\cite{cordis},
considering all the R\&D collaborations in the nine European Framework
Programmes, from 1984 until December 2025. For each of them, we
have information about the projects (ID and description) and the
organizations (name, country, city, and assigned funding) that work on
them. We obtain a temporal dataset from which we build the collaboration
network in which new countries enter over time. For a European focus, we
consider here only the 27 EU countries, 4 EFTA (European Free Trade
Association) countries, and the UK, for a total of 32 countries.

Our first aim is to investigate how inequality and integration change in
the European research collaboration network, understanding how less
privileged actors may acquire importance over time. The network, in
fact, consists of only 17 countries at the beginning, and is populated
and expanded over time in the following programmes. More peripheral and
less populated countries, or countries joining the EU later may start in
a disadvantaged position, participating in fewer projects and receiving
less funding, but what happens after some time? 

To assess how inequality and integration evolve within the European research collaboration system, we analyze the structure of collaborations from a country-level perspective. Specifically, we construct, for each FP, a collaboration network in which countries are connected through their participation in joint projects. This representation allows us to capture both the distribution of participation across countries and the overall cohesiveness of the system. We then quantify inequality by examining how projects are distributed across countries relative to their population size, and integration by measuring the efficiency of information flow within the network using Global Communication Efficiency~\cite{bertagnolliQuantifyingEfficientInformation2021}. Together, these metrics provide a complementary view on whether the expansion of the collaboration system is associated with more balanced participation and stronger systemic connectivity.

The second part of the work, instead, focuses on the thematic areas of the European research and on its exploration and expansion dynamics. 
We therefore analyze the semantic embedding space of the descriptions of the projects, focusing on 
the topics that characterize it and on its structure.
We identify macro-areas of interest through a hierarchical approach and we observe the shape,
evolution, and expansion of the space.
We are interested in which direction the European research moves, whether it
focuses on certain areas without exploring others, or it is driven by
innovation and exploration of new frontiers, and if and when research
interests change or expand. To evaluate these dynamics, we compute the
minimum spanning tree length on points (projects) over time and then compare it to a
null model. 
Finally, we perform a comparison between research interests and funding
in research-related and industry-related organizations. We are
interested in understanding to what extent companies, in different
countries, are involved in European projects compared to universities
and research centres, and in which fields.

Our research questions can be summarized as follows.

\noindent\textbf{(RQ1)} Does the progressive expansion of the European
collaboration network reduce structural inequalities in research
participation, and how does integration evolve for both incumbent and
newly entering countries?

\noindent\textbf{(RQ2)} Does the thematic landscape of European
research exhibit genuine exploratory diversification over time, or does
it follow a path-dependent trajectory of focused concentration?

\noindent\textbf{(RQ3)} How does the balance between research and
industry funding vary across thematic domains and countries, and what
does this reveal about the alignment between public research priorities
and private sector engagement in the European innovation ecosystem?

By answering them, we try to give an overview and explanation of some of
the main dynamics that characterize the European research landscape,
focusing on how countries tend to collaborate and towards which topics
there is more attention over time. In doing so, we contribute to the
science and innovation policy literature by providing the first
integrated longitudinal analysis that jointly examines the structural
evolution of the European collaboration network and the semantic
evolution of its research portfolio across all nine FPs.

The remainder of the paper is organized as follows. Section~\ref{sec:rel_work} reviews
the relevant literature and positions our contribution. Section~\ref{sec:data}
describes the data. Section~\ref{sec:collaboration_network} analyzes the European collaboration
network. Section~\ref{sec:semantic_space} examines the semantic space of research topics.
Section~\ref{sec:discussion} discusses the results and their policy implications.
Section~\ref{sec:conclusion} concludes.

\section{Background and Related Work}
\label{sec:rel_work}
Our study draws on and contributes to three strands of literature: the
analysis of European research collaboration networks, the geography and
inequality of science funding, and the thematic evolution of research
portfolios.

The collaborative structure of the European Framework Programmes has
attracted substantial scholarly attention since the early 2000s.
Roediger-Schluga and Barber~\cite{roediger2008} provided one of the
first systematic analyses of R\&D collaboration networks in the FPs,
documenting the emergence of a dense core of highly connected
organizations and countries. Breschi and Cusmano~\cite{breschi2004}
extended this analysis by examining how network structure relates to
knowledge flows and innovation performance, finding that participation
in FP networks is highly concentrated among a small number of
institutions in a few countries. Subsequent work has examined how the
network evolves with successive programme
generations~\cite{barber2008}, noting that while the overall density
of collaboration increases, the relative positions of countries remain
remarkably stable.

The gravity model has been widely applied to understand the determinants
of scientific collaboration~\cite{katz1994,ponds2007}. Hoekman,
Frenken, and Tijssen~\cite{hoekman2010} showed that geographical
distance remains a significant predictor of collaboration intensity in
European research, even after controlling for institutional and cultural
factors. Frenken, Hardeman, and Hoekman~\cite{frenken2009} further
demonstrated that whereas distance effects have diminished over time in
some scientific fields, they have not disappeared, suggesting that
proximity still matters for the formation of research partnerships. Our
use of a gravity model to track the evolving role of distance across all
nine FPs extends this line of inquiry over a substantially longer time
horizon.

A central tension in EU research policy is the trade-off between
excellence and cohesion~\cite{doussineau2020,morano2005}. FPs have historically concentrated their funding in countries
with stronger research systems, raising concerns about a ``Matthew
effect'' in which already-advantaged countries capture
disproportionate shares of resources~\cite{merton1968}.
Enger~\cite{enger2018} documented persistent geographical disparities
in FP participation, with Southern and Eastern European countries
systematically underrepresented relative to their scientific capacity.
The introduction of ``widening'' instruments in Horizon 2020 such as
Teaming, Twinning, and ERA Chairs represented a policy response to these
disparities, though evidence on their effectiveness remains
mixed~\cite{fresco2015}.

Our analysis of inequality (via the Gini index) and integration (via
Global Communication Efficiency) across all nine FPs provides a more
comprehensive temporal perspective on this debate than has been
available to date. By decomposing these metrics across successive
cohorts of countries entering the network, we can disentangle the
effects of network expansion from the dynamics of progressive
integration.

The evolution of research topics over time has been explored through a range of computational approaches. Previous studies have examined how researchers move across topics and disciplines, conceptualized as a ``diaspora of knowledge''~\cite{de2016quantifying}, while topic modeling techniques, from LDA to more recent transformer-based methods such as BERTopic~\cite{grootendorstBERTopicNeuralTopic2022}, have been used to map the thematic structure of scientific production~\cite{blei2003}. However, this literature has largely focused on individual trajectories or specific domains, with limited attention to the aggregate evolution of research priorities within large-scale funding systems.

Our approach to the characterization of the semantic space of European
research through minimum spanning tree (MST) analysis in embedding space
is, to our knowledge, novel in this context. By comparing the observed
MST length to a null model of random exploration, we can quantify the
degree to which European research exhibits focused exploitation versus
broad exploration: a distinction with direct relevance to debates about
mission-oriented innovation policy~\cite{mazzucato2018} and the
capacity of FPs to support transformative
research~\cite{kuhlmann2018}.

\deleted{}

\section{Data}
\label{sec:data}

In this section, we present the data leveraged for this study (Section \ref{sec:data_overview}) and a first exploratory data analysis (Section \ref{sec:exp_analysis}).

\subsection{Data overview}
\label{sec:data_overview}
Our study relies on publicly available data from the CORDIS database \footnote{\url{https://cordis.europa.eu/}} \footnote{\url{https://data.europa.eu/data/datasets?locale=en}}, covering R\&D collaborations across the nine FPs. Data include project-level information (IDs and descriptions) and organization-level attributes (name, country, city, and funding).

We thus have two types of datasets for each FP, the
\textit{Projects}, which presents all the projects in the corresponding
FP, and the \textit{Organizations}, which reports all the
project-organization pairs, matching the organizations that took part in
any specific project. In the \textit{Projects} dataset, in particular,
we are interested in the fields \textit{id}, which contains the IDs of
the projects, \textit{startDate} and \textit{endDate}, which report
the start and end dates of the projects, and \textit{objective}, which
contains the short descriptions (abstracts) of the projects.

In the \textit{Organizations} dataset, instead, we are interested in
the fields \textit{projectID} (projects IDs corresponding to the ones
in \textit{Projects}), \textit{organisationID} (organizations IDs),
\textit{name} (organizations names), \textit{activityType} (type of
organization), \textit{city}, \textit{country}, \textit{geolocation}
(related to the organizations' location), \textit{netEcContribution}
(funding assigned to a specific organization on a specific project),
\textit{role} (role of the organization in the specific project, in
particular whether they served as coordinator or participant).

Organizations, in \textit{activityType}, are classified as
\textit{Higher or Secondary Education Establishments (HES)},
\textit{Research Organizations (REC)}, \textit{Private for-profit
entities (PRC)}, \textit{Public Bodies (PUB)}, \textit{Other (OTH)}.
We group them in three sets, according to whether they are more
research-oriented or industry-oriented, or neither specifically. The
\textit{Research} set includes \textit{HES} and \textit{REC}, the
\textit{Industry} set corresponds to \textit{PRC}, and the
\textit{Other} set includes \textit{PUB} and \textit{OTH}. In this
study, we work at a country level and focus on a subset of 32 European
countries: 27 countries from the European Union, 4 EFTA (European Free
Trade Association) countries, and the United Kingdom. The full list of the 32 countries with their ISO2 code can be found in Appendix \ref{app:table_countries}. 
A detailed overview of the data can be found in
Table~\ref{tab:data}. As shown, there are nine Framework Programmes:
the first started in 1984 and the latest, the current Horizon Europe, began in 2021. 
The table reports an overview of our
data, specifying, for each FP, years, number of
projects, countries involved, funding (when available), and number of
unique organizations (when available).
The number of
funded projects increases over time and also the number of countries
taking part in them.
It is worth noticing that the Horizon Europe programme ends in 2027 (as
shown in Table~\ref{tab:data}). Here, we focus on the projects
starting before January 2026, as the programme is still ongoing while we
perform this study.

Among our 32 countries, some are present since the beginning, whereas
others become part of the research programmes over time. Only from FP4
on, all 32 countries are present. The CORDIS database has been populated
over time and, for this reason, presents some inconsistencies in the
format of the data and some missing information. Focusing on the
features that match our specific research interests, funding is
consistently reported only in the last two FPs, H2020 and Horizon
Europe, and the organizations present a unique identifier only from FP7
onward. The absence of unique identifiers before FP7, together with
inconsistencies in organization naming and missing geolocation data,
required substantial data harmonization. We therefore created a pipeline for organizations cities retrieval that leverages the combination of different tools and approaches (i.e. Nominatim \cite{Nominatim} as geocoding system, Levenshtein distance \cite{levenshtein1966binary}, Google search, and LLM). The details of the pipeline are described in Appendix \ref{app:geo_city}. 
In this study, however, we carry out our analysis mainly at the country
level, avoiding possible issues due to these inconsistencies.

\begin{table}
\begin{center}
\begin{tabular}{ p{1.5 cm} p{1.9 cm} p{1.7 cm} p{1.9 cm} p{2.9 cm}
p{1.7 cm} p{2.7 cm}}
\hline
\textbf{FP} & \textbf{Years} & \textbf{Projects} &
\textbf{Countries} & \textbf{EU+EFTA+ UK countries} & \textbf{Funding
(billions euros)} & \textbf{Organizations}\\
\hline
\textbf{FP1} & 1984-1987 & 3282 & 26 & 17 & NA & NA \\
\textbf{FP2} & 1987-1991 & 3884 & 68 & 19 & NA & NA \\
\textbf{FP3} & 1990-1994 & 5527 & 112 & 27 & NA & NA \\
\textbf{FP4} & 1994-1998 & 14526 & 145 & 32 & NA & NA \\
\textbf{FP5} & 1998-2002 & 17206 & 150 & 32 & NA & NA \\
\textbf{FP6} & 2002-2006 & 10093 & 157 & 32 & NA & NA \\
\textbf{FP7} & 2007-2013 & 25785 & 178 & 32 & NA & 30566 \\
\textbf{H2020} & 2014-2020 & 35389 & 180 & 32 & 68.3 & 41824 \\
\textbf{Horizon Europe} & 2021-2025 & 17521 & 178 & 32 &
48.3 & 29346 \\
\hline
\end{tabular}
\end{center}
\caption{\textbf{Overview of the data.} The table reports information
about the nine datasets related to the nine FPs. In
particular, we report, for each FP, years of operation, number of
projects, number of participating countries (total and belonging to the
set EU+EFTA+UK), funding (when available), number of organizations (when
available). To notice, for Horizon Europe, we report the data until December 2025 even though the programme is still ongoing while conducting this study and will end in 2027.}
\label{tab:data}
\end{table}

\subsection{Exploratory data analysis}
\label{sec:exp_analysis}

We show here an exploratory analysis of our data, focusing on the
distribution of projects and funding per geographical areas in H2020,
which is the largest and most complete dataset and for which we have both the information
about funding amounts and number of projects.

\begin{figure}[H]
\centering
\includegraphics[width=1\textwidth]{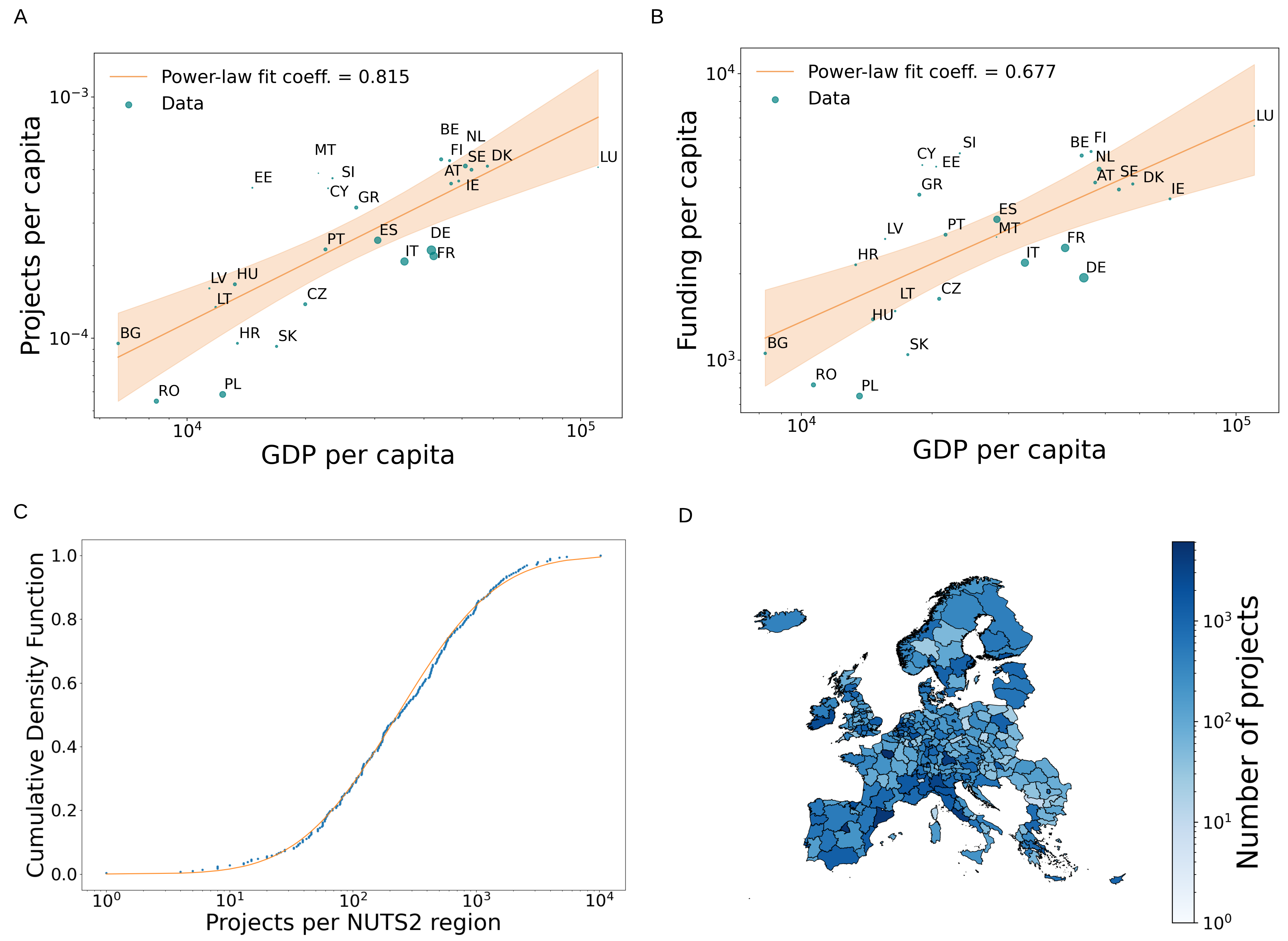}
\caption{\textbf{Exploratory data analysis.} \textbf{(A)} GDP
per capita VS projects per capita in log-log scale in H2020 on the 27 EU
countries: points sized by countries' population, Spearman's rank
correlation coefficient of 0.68. \textbf{(B)} GDP per capita VS funding
per capita in log-log scale in H2020 on the 27 EU countries: points
sized by countries' population, Spearman's rank correlation
coefficient of 0.67. \textbf{(C)} Cumulative density function of number
of projects per NUTS2 region in H2020 on 32 countries (EU + EFTA + UK):
data appears to follow a lognormal distribution, with a Gini index of 0.700. \textbf{(D)} Number of projects
per NUTS2 region in H2020 on 32 countries (EU + EFTA + UK):
areas with major concentration are \^Ile-de-France, Comunidad de
Madrid, Catalu\~na, Oberbayern, and Lazio.}
\label{fig:geography}
\end{figure}

In Figure~\ref{fig:geography}A we observe a positive correlation
between GDP per capita per country and projects per capita per country
with a Spearman's rank correlation coefficient of 0.68
(\textit{p}-value 0.0001), whereas in Figure~\ref{fig:geography}B we
find a positive correlation between GDP per capita per country and
funding per capita per country with a Spearman's rank correlation
coefficient of 0.67 (\textit{p}-value 0.0001). In both cases we report
the data from H2020 for the 27 EU countries.

These correlations are consistent with the well-documented ``Matthew
effect'' in European research funding~\cite{merton1968}, whereby
countries with already stronger economies and research systems capture
disproportionately more projects and funding. Importantly, the power-law
exponents differ between the two relationships (0.815 for projects,
0.677 for funding), suggesting that while wealthier countries
participate in more projects, the funding advantage is less steep,
possibly reflecting policy mechanisms that partially mitigate economic
disparities in per-project allocation.

Figure~\ref{fig:geography}C shows, through the blue points, the cumulative density function of
the number of projects per NUTS2\footnote{\url{https://ec.europa.eu/eurostat/web/nuts/history}} region (level 2 of Nomenclature of Territorial Units for Statistics, which correspond to basic, intermediate-level territorial units) of the 32
countries (EU + EFTA + UK) in H2020: we observe that it resembles 
the cumulative density function of a lognormal distribution (orange
line). The lognormal is a long-tailed distribution function, typically associated with high inequality. This is consistent with the European research funding distribution, where countries with high stickiness and attractiveness scores show a rich-get-richer effect in winning funds, as observed in \cite{de2016eu}.

A similar behavior in projects participation is also suggested by the distribution of projects per NUTS2. 
For a lognormal distribution $X \sim \mathrm{LogNormal}(\mu, \sigma^2)$, the Gini coefficient depends only on the scale parameter (standard deviation) $\sigma$ and is given by $G = \operatorname{erf}\!\left(\frac{\sigma}{2}\right)$, with $\operatorname{erf}$ the error function. Given $\sigma = 1.456$ from our lognormal fit, we observe a Gini index $G = 0.700$, indicating, as expected, a highly inhomogeneous distribution of projects.

In Figure~\ref{fig:geography}D, instead, we report the
number of projects per NUTS2 region. In particular, we can observe that
there are some regions with a larger number of participations to
projects, such as \^Ile-de-France in France, Comunidad de Madrid and
Catalu\~na in Spain, Oberbayern in Germany, and Lazio in Italy. The
headquarters of the major national research institutes of these
countries are located in such regions, which partly explains the
observed concentration.


\section{European collaboration network}
\label{sec:collaboration_network}

\begin{figure}[H]
\centering
\includegraphics[width=1\textwidth]{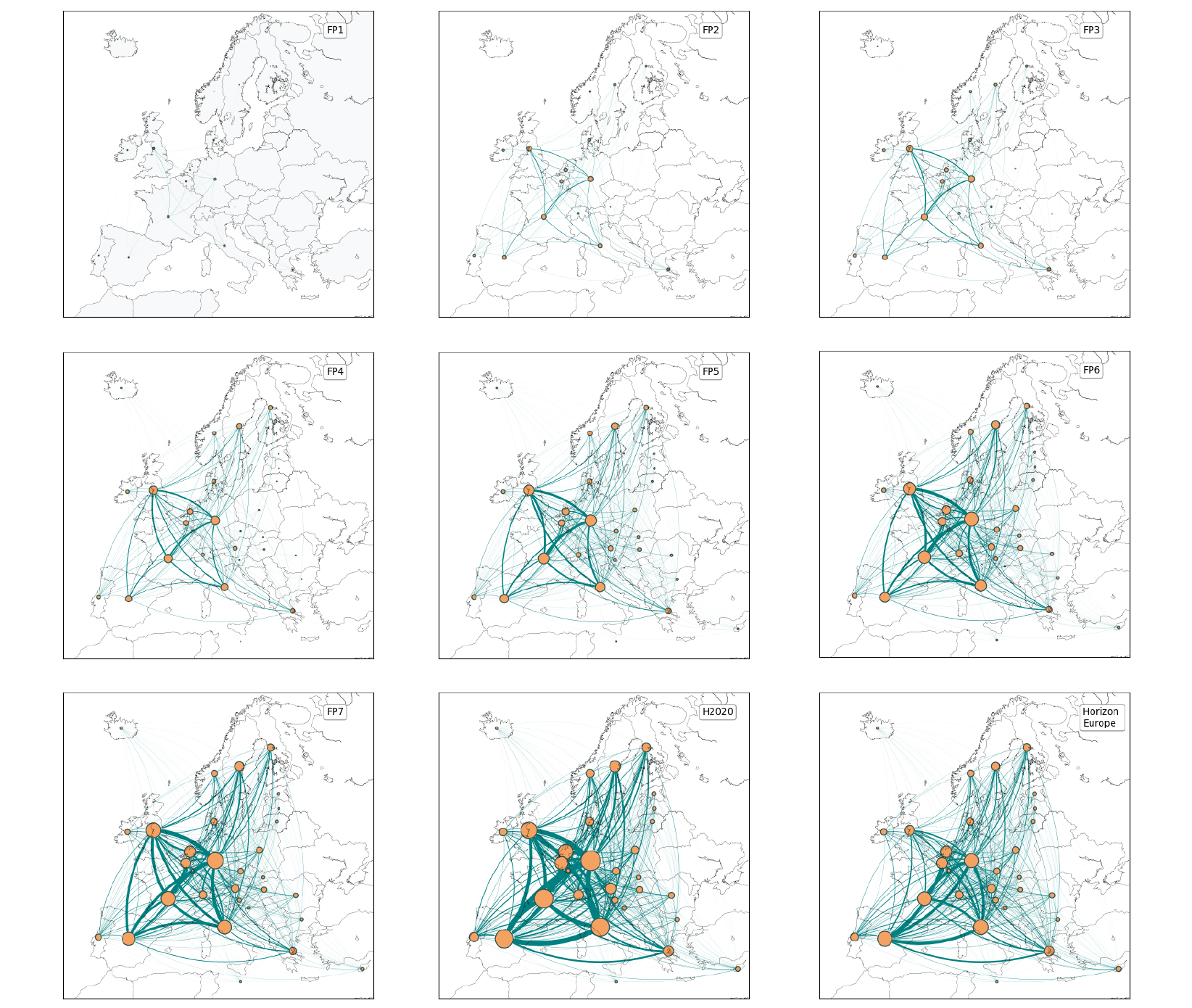}
\caption{\textbf{Evolution of the European collaboration network.} The panel shows, for each FP, the collaboration network between the countries taking part into projects. Networks are undirected and weighted, nodes are sized by strength and edges by the number of collaborations between countries. New countries enter the network over time, the number of projects and collaborations increases over time. The last network (Horizon Europe) considers only the collaborations on projects starting before the end of 2025. Data for the boundaries was extracted from the GSHHG dataset \cite{GSHHG}.}
\label{fig:networks}
\end{figure}

For each FP, we aggregate inter-organizational collaborations at the country level, yielding an undirected weighted network whose nodes represent countries and whose edge weights capture the number of joint projects between each pair. Figure~\ref{fig:networks} displays the resulting networks across the nine FPs. The collaborative space expands progressively: from 17 countries in FP1 to 19 in FP2, 27 in FP3, and the full set of 32 from FP4 onward. In parallel, as discussed in Section~\ref{sec:data}, both the number of projects and participating organizations grow steadily, producing increasingly dense collaborative structures.

\subsection{How does distance influence collaborations?}

To assess the role of geographical proximity in shaping collaboration patterns, we fit a gravity model that predicts the number of joint projects between each pair of countries on the basis of three variables: population, GDP per capita, and geographical distance. Gravity models have been widely adopted in the science-of-science literature to characterize the determinants of research collaboration~\cite{katz1994,ponds2007,hoekman2010}; our application extends this approach over a substantially longer time horizon, covering all nine FPs.

We denote by $\Phi_{XY}$ the estimated number of collaborations between countries $X$ and $Y$. The model incorporates the population of each country ($p_X$, $p_Y$), their GDP per capita ($g_X$, $g_Y$), and the geographical distance between their centroids ($r_{XY}$):

\begin{equation}
\Phi_{XY} = k \frac{p_{X}^{\alpha_1} p_{Y}^{\alpha_2}
g_{X}^{\beta_1} g_{Y}^{\beta_2}}{r_{XY}^\gamma}
\end{equation}

where $k$ is a normalization factor and $\alpha_1$, $\alpha_2$, $\beta_1$, $\beta_2$, $\gamma$ are fitted coefficients.

\begin{figure}[H]
\centering
\includegraphics[width=0.9\textwidth]{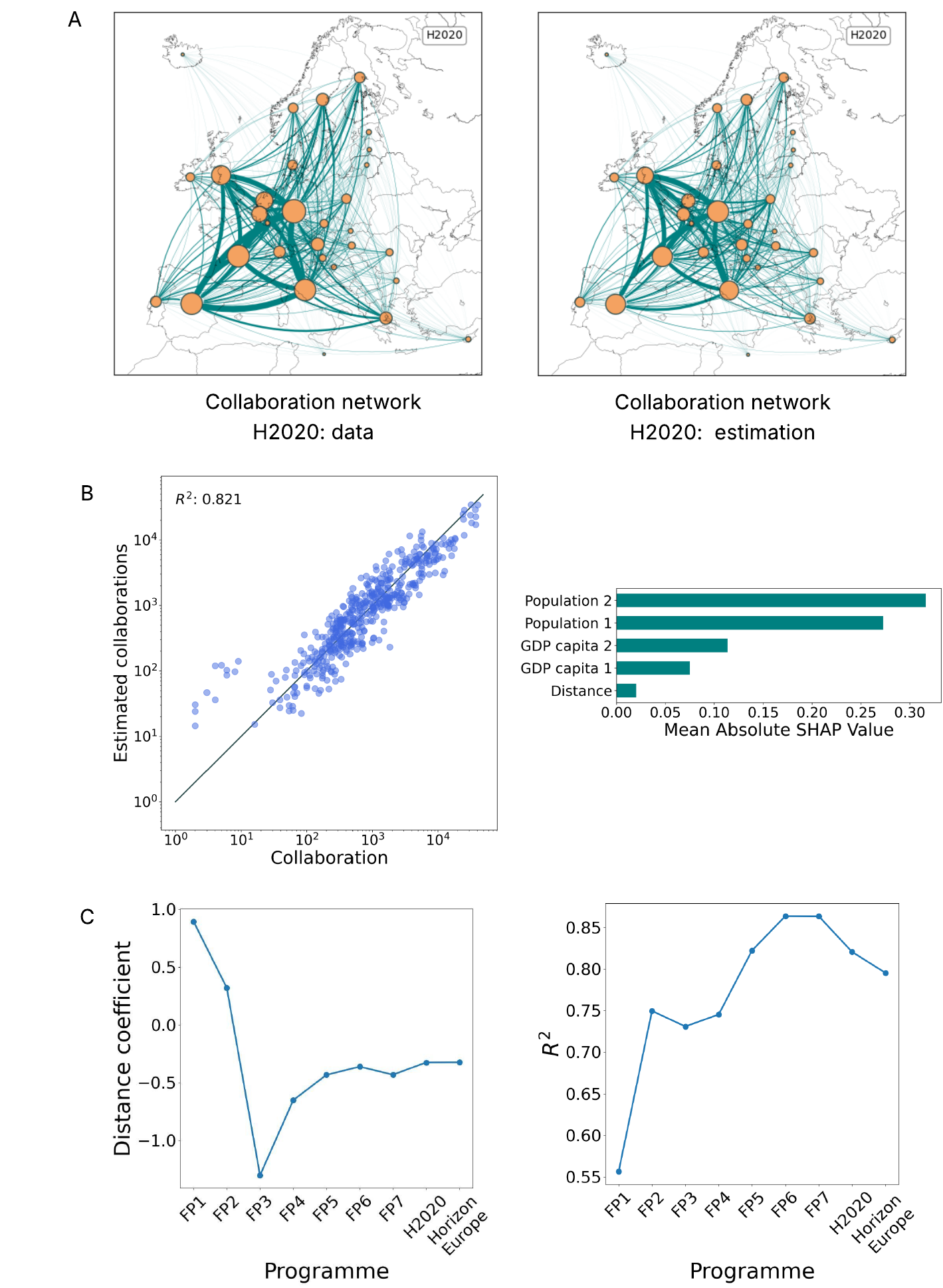}
\caption{\textbf{Gravity model: estimating collaborations between countries.}
\textbf{(A)} The collaboration network of H2020: on the left the one built from the data; on the right the one estimated through the gravity model. \textbf{(B)} Collaborations in H2020 from data VS collaborations in H2020 estimated through the gravity model on the left and the importance of the predictors through SHAP on the right. \textbf{(C)} The evolution of the distance coefficient (left) and of $R^2$ (right).}
\label{fig:gravity}
\end{figure}

Figure~\ref{fig:gravity} summarizes the results. Panel~\ref{fig:gravity}A compares the observed H2020 collaboration network with its model estimate. Panel~\ref{fig:gravity}B plots observed against predicted collaboration counts for H2020, alongside predictor importances computed via SHAP values. Panel~\ref{fig:gravity}C traces the evolution of the distance coefficient $\gamma$ and the model fit $R^2$ across the nine programmes.

The gravity model achieves a good fit, particularly in the later programmes, with $R^2$ stabilizing between 0.8 and 0.9 from FP5 onward. The few outlier points in the lower-left region of Figure~\ref{fig:gravity}B, where collaboration counts are substantially overestimated, correspond to partnerships involving Liechtenstein, a country with the highest GDP per capita in our sample but very limited participation in FP projects, leading the model to overpredict its collaborative output. The evolution of the distance coefficient $\gamma$ reveals three distinct regimes. In the early programmes, $\gamma$ is positive, indicating that geographical proximity favours collaboration. In FP3, $\gamma$ drops sharply to a negative value, implying that collaborations during this period were disproportionately long-distance, plausibly reflecting the establishment of new partnerships with the many countries entering the network at that time. From FP4 onward, $\gamma$ stabilizes closer to zero, but still negative, indicating that distance has become a more marginal factor in determining collaboration intensity, but still favours long-distance relationships. The SHAP analysis corroborates this picture: population (country size) emerges as the most influential predictor, followed by GDP per capita, with distance ranking last.

This finding extends the results of Hoekman et al.~\cite{hoekman2010}, who documented a declining but non-negligible role of distance in European scientific collaboration. Our longer temporal perspective reveals that by the later FPs, geographical distance has effectively ceased to be a significant determinant of collaboration patterns, with country size and economic capacity playing substantially larger roles. The negative peak in $\gamma$ during FP3, the programme during which the largest number of new countries entered the network, may reflect a deliberate policy effort to establish long-distance partnerships with newly integrated members, a pattern that subsequently normalizes as these partnerships become routine.

\subsection{Inequality and integration}

We next examine how the progressive expansion of the network affects the distribution of participation across countries. We operationalize this question through two complementary metrics: inequality, measured by the Gini index on the distribution of projects per capita per country, and integration, quantified via Global Communication Efficiency~\cite{bertagnolliQuantifyingEfficientInformation2021}, which captures the efficiency of the information flow through the network as determined by its topology.

\begin{figure}[H]
\centering
\includegraphics[width=1\textwidth]{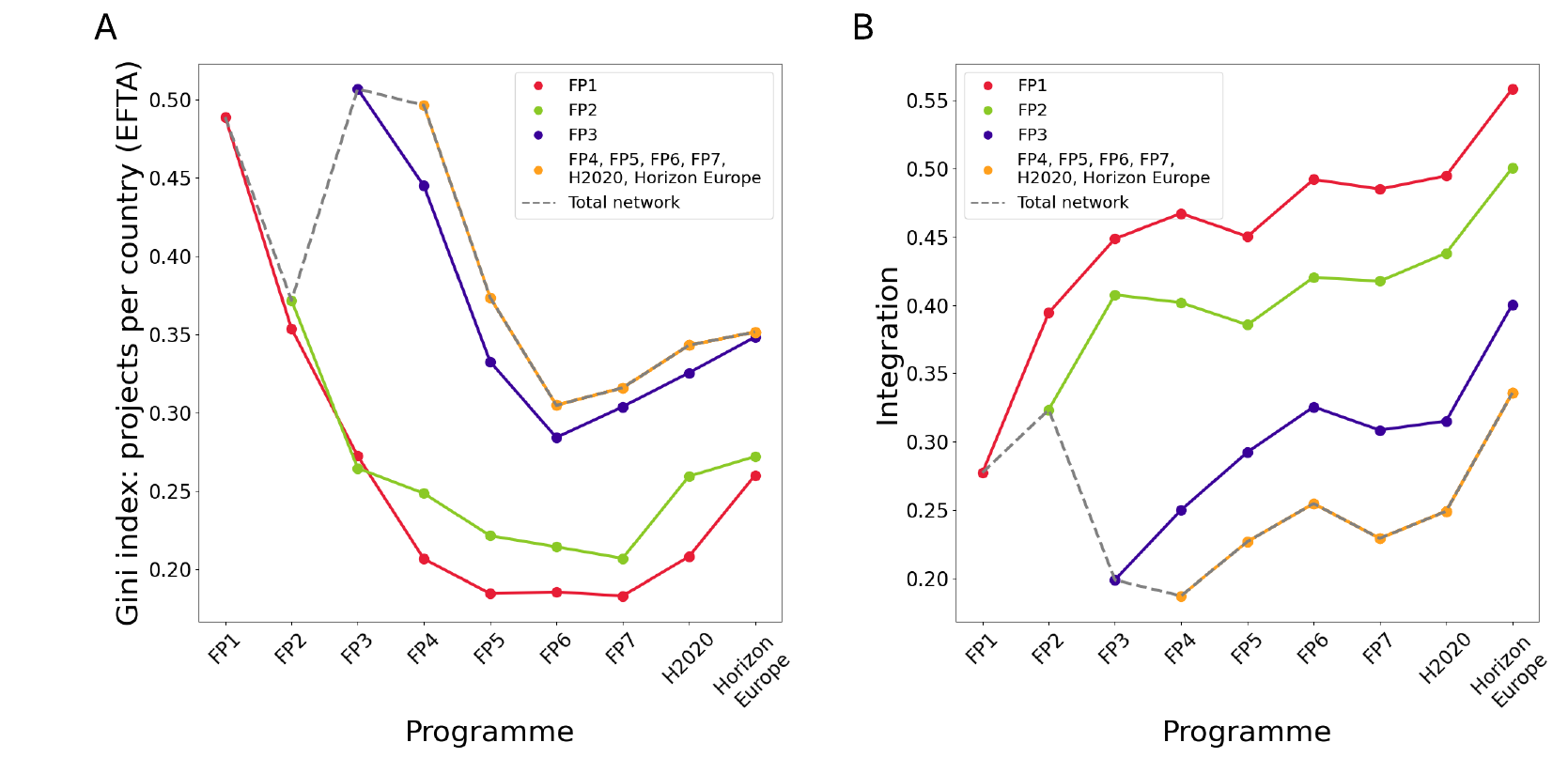}
\caption{\textbf{Evolution of inequality (Gini index) and integration (Global Communication Efficiency) over time.} \textbf{(A)} The dashed grey line indicates the Gini index on the distribution of the total number of projects per capita per country, computed for each of the nine FPs. Inequality decreases and reaches the minimum in FP6, whereas it has a slight increase in the last FPs. For the other coloured lines, only the subset of countries appearing in the corresponding programme (as detailed in the legend) is used to create the subsets on which inequality is computed. Inequality tends to be higher when adding more countries to the sets, but if we consider each specific subset, it tends to decrease over time and to have a slight increase in the last programmes.
\textbf{(B)} The dashed grey line indicates the integration computed through Global Communication Efficiency on the total network of countries participating in each FP. For the other lines, only the countries appearing in the corresponding programme (as detailed in the legend) are used to create the subnetworks on which integration is computed. By adding new nodes (from the red line with 17 countries to the purple line with 32 countries) integration decreases, but nodes already in the network become more integrated in the following programmes.}
\label{fig:ineq_integr}
\end{figure}

Figure~\ref{fig:ineq_integr}A displays the evolution of inequality. The dashed grey line, computed over all countries participating in each FP, peaks during FP3 and FP4, the programmes during which most new countries join the network. From FP4 onward, when all 32 countries are present, inequality declines and stabilizes between 0.30 and 0.35, with a modest upward trend in the most recent programmes. To disentangle the effects of network expansion from within-cohort dynamics, we decompose the analysis by country cohort (coloured lines in Figure~\ref{fig:ineq_integr}A). Each line tracks the Gini index computed over a fixed subset of countries: the 17 of FP1 (red), the 19 of FP2 (green), the 27 of FP3 (purple), and the full 32 from FP4 onward (yellow). A consistent pattern emerges: adding new countries raises inequality, but within each cohort the Gini index drops markedly in subsequent programmes, before exhibiting a slight increase in the most recent ones.

Figure~\ref{fig:ineq_integr}B reports the corresponding evolution of integration. The dashed grey line shows that overall integration reaches its minimum during FP3 and FP4, precisely when the largest influx of new countries occurs. From FP4 onward, integration increases steadily. The cohort decomposition reveals the underlying mechanism: when new nodes enter the network, integration temporarily decreases, but these nodes become progressively better integrated in subsequent programmes, as evidenced by the upward trend of each cohort line.

Together, these results paint a nuanced picture of European research integration. The within-cohort decline in inequality, combined with a slight aggregate upward trend in recent programmes, echoes concerns raised in the ``widening'' debate~\cite{ec_widening,fresco2015}: while new entrants are progressively absorbed into the network, the most recent programme generations may be introducing new sources of disparity, possibly driven by the increasing emphasis on excellence-based competitive funding~\cite{doussineau2020}. The integration dynamics offer a more encouraging reading: once countries enter the network, they consistently become more integrated over successive programmes, suggesting that the collaborative infrastructure of FPs does fulfil its intended function as an instrument of progressive convergence, even if the pace remains uneven.

\section{Semantic space: evolution of research interests}
\label{sec:semantic_space}

The second part of this study turns from the structure of collaborations to the content of research itself. We pursue three related objectives: identifying the thematic macro-areas on which European research concentrates, characterizing how the semantic space of research expands over time, and comparing funding allocations between research organizations and industry across topics.

\subsection{Semantic embeddings extraction and topic modeling}
\label{sec:embeddings}

We extract vector representations from the \textit{objective} field of our \textit{Projects} dataset, which contains the project abstract, using the pretrained model \texttt{BGE-M3}\footnote{\url{https://huggingface.co/BAAI/bge-m3}}, a multilingual encoder that maps text to a 1024-dimensional dense vector space~\cite{chenM3EmbeddingMultiLingualityMultiFunctionality2025}. We retain only projects with a non-empty \textit{objective} field, yielding a corpus of 122,570 documents. A key advantage of \texttt{BGE-M3} is its support for input sequences of up to 8,192 tokens, accommodating project descriptions that exceed the 512-token limit of many alternative models.

We perform topic modeling with BERTopic~\cite{grootendorstBERTopicNeuralTopic2022}, which combines transformer-based embeddings with clustering. The pipeline first reduces the embeddings to two dimensions via UMAP~\cite{mcinnes2018umap}, for which we keep the BERTopic's default value of the parameter that controls for the more global or local view of the embedding structure (\texttt{n\_neighbors}$\,{=}\,$15), while we empirically set the reduced dimensionality to 2 (\texttt{n\_components}$\,{=}\,$2), consistent with the dimensionality needed for visualization. The second step of the pipeline consists of clustering the embeddings with HDBSCAN~\cite{campello2013density}, for which we calibrate the parameter that controls for the minimum size of a cluster and consequently of the number of generated clusters to the dataset size to avoid excessively broad or narrow topics (\texttt{min\_cluster\_size}$\,{=}\,$100). This configuration yields 117 initial topics.

HDBSCAN typically leaves a fraction of documents unassigned. We reduce these outliers using BERTopic's c-TF-IDF strategy\footnote{\url{https://maartengr.github.io/BERTopic/getting_started/outlier_reduction/outlier_reduction.html}}\footnote{\url{https://maartengr.github.io/BERTopic/getting_started/ctfidf/ctfidf.html}}, which computes c-TF-IDF representations for outlier documents and assigns each to the best-matching non-outlier topic, so that every document is ultimately associated with a topic.

We then apply hierarchical topic modeling to merge the 117 fine-grained topics into broader research areas. BERTopic uses SciPy's \texttt{ward} linkage function\footnote{\url{https://maartengr.github.io/BERTopic/getting_started/hierarchicaltopics/hierarchicaltopics.html}}\footnote{\url{https://docs.scipy.org/doc/scipy/reference/generated/scipy.cluster.hierarchy.linkage.html}} by default, which minimizes within-cluster variance and produces compact, well-separated hierarchies. We examine the resulting dendrogram and identify 16 macro-clusters by combining a distance threshold with manual verification of thematic coherence. We assign names to these broad macro-areas by inspecting the representative keywords of the merged topics.

Some documents may be less accurately classified due to multidisciplinary content or generic descriptions. We acknowledge that alternative approaches allowing probabilistic topic assignment, such as considering the top-$k$ topics by probability for each document, could be more appropriate for analyses requiring finer-grained thematic resolution, but the macro-level granularity adopted here is consistent with our research questions.

\begin{figure}[H]
\centering
\includegraphics[trim=5cm 1.5cm 0 0, clip, width=1\textwidth]{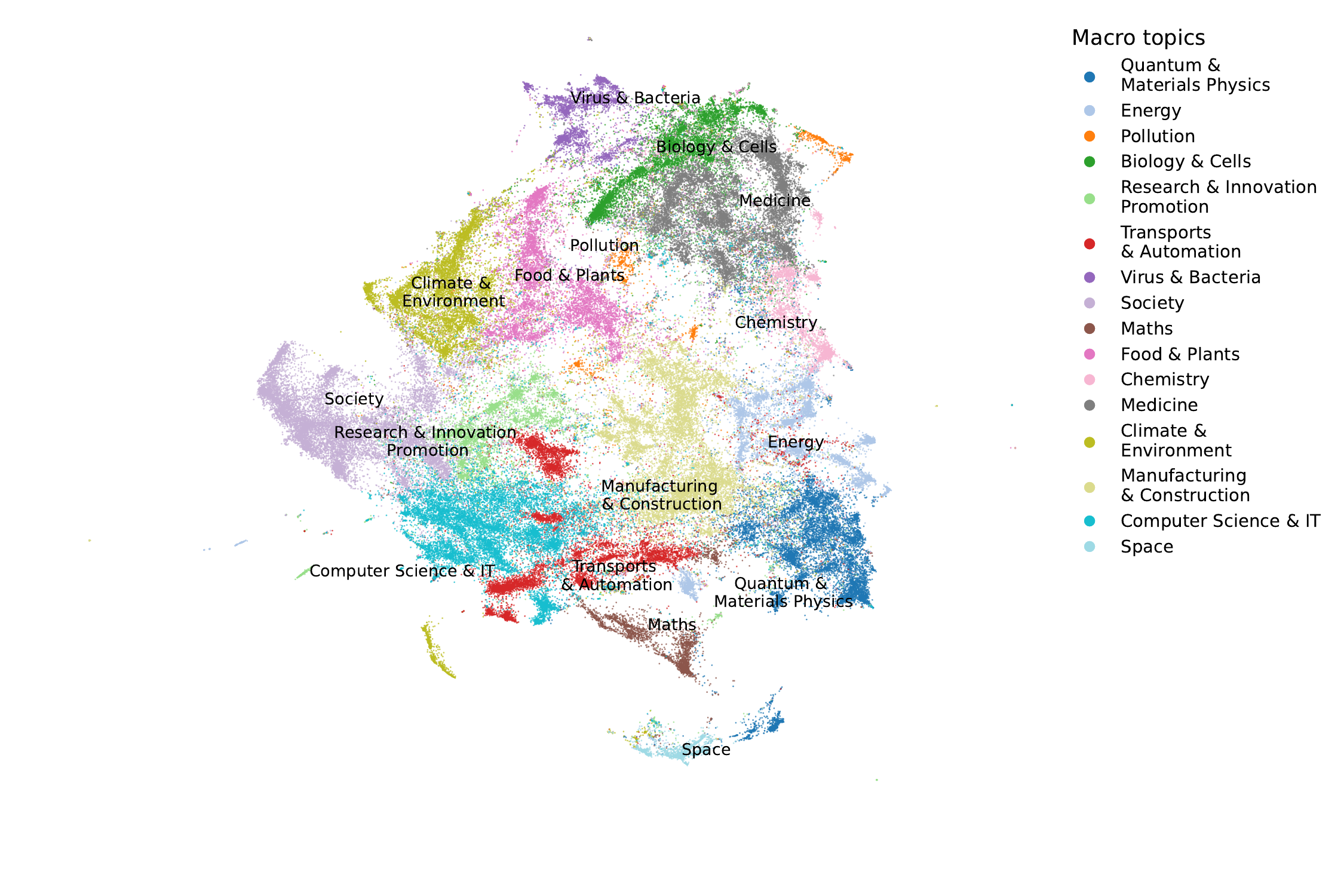}
\caption{\textbf{Embedding space and macro-topics.}
The embeddings of the project descriptions coloured by the 16 macro-topics obtained from the merge of the initial 117 topics are shown.}
\label{fig:emb_space}
\end{figure}

Figure~\ref{fig:emb_space} displays the embedding space coloured by the 16 macro-topics. The spatial arrangement is thematically coherent: biomedical topics, for instance, cluster together in the top-right corner of the space, reflecting their semantic proximity.

\begin{table}
\begin{center}
\begin{tabular}{ p{4.8 cm} p{2.5 cm} p{3.5 cm} p{3.5 cm} }
\hline
\rule{0pt}{7ex}\textbf{Macro-topic} & \textbf{\shortstack[l]{Total \\ projects}} & \textbf{\shortstack[l]{Projects H2020 \\ \& HEU}} &
\textbf{\shortstack[l]{Funding \\ H2020 \& HEU \\ (billions euros)}} \\[4 pt]
\hline
\rule{0pt}{4ex}\textbf{Society} & 15109 & 8384 & 4.86 \\[5 pt]
\textbf{\shortstack[l]{Manufacturing \& \\Construction}} & 12247 & 3762 & 1.70 \\[5 pt]
\textbf{\shortstack[l]{Computer Science \& \\ IT}} & 11893 & 4195 & 1.70 \\[5 pt]
\textbf{Medicine} & 11636 & 6708 & 4.41 \\[4 pt]
\textbf{\shortstack[l]{Quantum \& \\ Materials Physics}} & 10257 & 4626 & 3.59 \\[5 pt]
\textbf{\shortstack[l]{Transports \& \\ Automation}} & 8801 & 4210 & 1.71 \\[5 pt]
\textbf{\shortstack[l]{Climate \& \\ Environment}} & 8633 & 3109 & 1.91 \\[5 pt]
\textbf{\shortstack[l]{Biology \& \\ Cells}} & 8128 & 3501 & 3.31 \\[5 pt]
\textbf{\shortstack[l]{Research \& \\ Innovation Promotion}} & 7389 & 3140 & 2.03 \\[5 pt]
\textbf{\shortstack[l]{Food \& \\ Plants}} & 7387 & 2491 & 1.23 \\[5 pt]
\textbf{\shortstack[l]{Energy}} & 7278 & 3138 & 1.72 \\[5 pt]
\textbf{\shortstack[l]{Virus \& \\ Bacteria}} & 3920 & 1708 & 1.50 \\[5 pt]
\textbf{Chemistry} & 3422 & 1351 & 1.11 \\[5 pt]
\textbf{Maths} & 3201 & 1322 & 1.25 \\[5 pt]
\textbf{Pollution} & 1577 & 444 & 0.20 \\[5 pt]
\textbf{Space} & 1307 & 662 & 0.72 \\[5 pt]

\hline
\end{tabular}
\end{center}
\caption{\textbf{Projects and funding per macro-topic.} The table reports the number of projects per macro-topic for all the FPs and specifically for H2020 and Horizon Europe (HEU). It also shows the funding per macro-topic in H2020 and Horizon Europe.}
\label{tab:macro-topic}
\end{table}

The macro-topics vary substantially in size, reflecting the distribution of European research activity across thematic domains. As reported in Table~\ref{tab:macro-topic}, the largest macro-topic is \textit{Society} (15,109 projects), followed by \textit{Manufacturing \& Construction} (12,247), \textit{Computer Science \& IT} (11,893), and \textit{Medicine} (11,636); the smallest is \textit{Space} (1,307 projects). The breadth of \textit{Society} partly reflects the aggregation of diverse humanistic disciplines into a single cluster, whereas the more granular subdivision of scientific topics spreads their projects across multiple macro-categories. The dominance of \textit{Society}, \textit{Manufacturing \& Construction}, and \textit{Computer Science \& IT} reflects the dual orientation of FPs toward societal challenges and industrial competitiveness. The comparatively smaller size of areas such as \textit{Space}, \textit{Maths}, and \textit{Quantum \& Materials Physics} should not be interpreted as indicating lower importance, but rather the more specialized nature of these research communities.

\subsection{Embedding space expansion and exploration}
\label{sec:mst}

To characterize the overall evolution of European research, we measure what distance we need to travel in the semantic space to reach all the projects that are embedded in it, as a way to measure its expansion. We perform this measure through the computation of the minimum spanning tree (MST) length over time.
Unlike other measures, such as the density of points in the space, the volume of the convex hull determined by the points, or the radius of the minimum enclosing ball, the MST takes into account the position of every single point, and its length thus better represents how spread out these points are in their n-dimensional space.
The measures mentioned above primarily capture the overall expansion or contraction of the region defined by the points and are more sensitive to outlier points compared to MST.

We therefore partition the temporal range into one-year windows and, for each window, consider all points in the embedding space corresponding to projects starting within that period (based on the \textit{startDate} field). We compute the MST length of these points and compare it to a null model in which the same number of points is uniformly sampled within the bounding box of the observed data. We additionally compute the cumulative MST length, spanning all projects from the beginning of FP1 up to the selected year, to capture the progressive expansion of the research space.

A note on scaling is in order. Since \texttt{BGE-M3} includes a final normalization layer, the embeddings are L2-normalized. We compute the MST length using Euclidean distance and apply the theoretical scaling of Beardwood et al.~\cite{beardwoodShortestPathMany1959} and Steele~\cite{steele1988growth}, dividing by $n^{(d-1)/d}$, where $n$ is the number of points and $d=1024$ is the embedding dimension. 
As $d \to \infty$, the exponent $(d-1)/d$ approaches 1, so that the scaling factor converges to $n$. In our setting, where $d$ is large, $(d-1)/d \approx 1$ and the normalization can therefore be well approximated by dividing the MST length by $n$.
This normalization corrects for the dependence of inter-point distances on both sample size and dimensionality, making the MST length comparable across time windows with different number of projects. Because the scaling is asymptotically valid for large $n$, we exclude the first three time windows, which contain fewer than 500 projects.

\begin{figure}[H]
\centering
\includegraphics[width=1\textwidth]{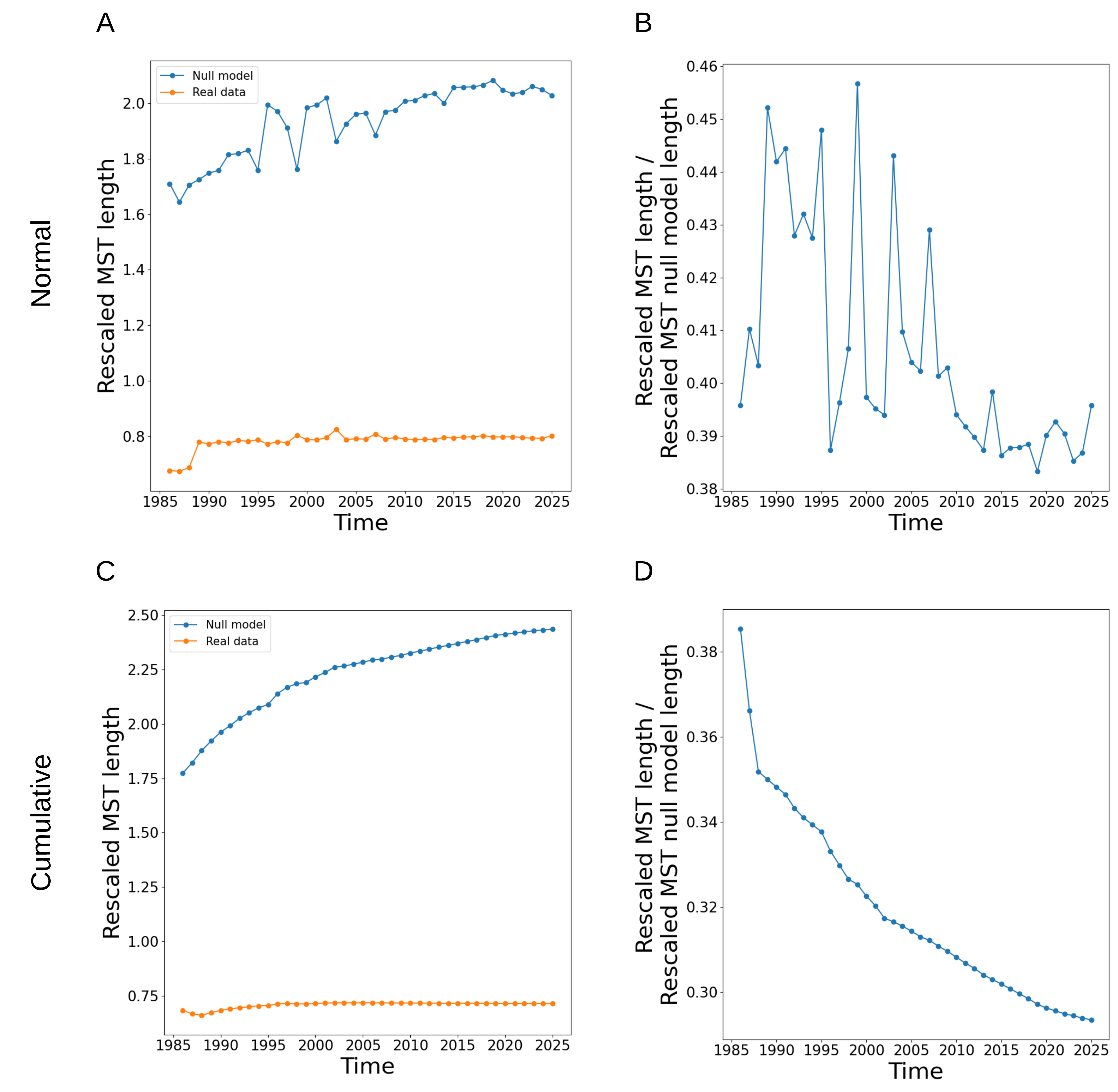}
\caption{\textbf{Temporal minimum spanning tree.} We report the computed rescaled MST length in different cases. The measure is obtained by dividing the MST length by $n^{(d-1)/d}$, where $n$ is the number of points and $d=1024$ is the embedding dimension.
\textbf{(A)} MST length computed on time windows of 1 year: data (in orange) VS null model (in blue). \textbf{(B)} Ratio between MST length of data and of null model in time windows of 1 year. \textbf{(C)} Cumulative MST length: data (in orange) VS null model (in blue). \textbf{(D)} Ratio between cumulative MST length of data and of null model.}
\label{fig:mst}
\end{figure}

Figure~\ref{fig:mst} presents the detailed results. Panel~\ref{fig:mst}A compares the MST length computed over yearly time windows in the data (orange) and in the null model (blue). The MST length in the data is consistently shorter than in the null model, indicating that European research does not explore the semantic space uniformly but instead concentrates in specific thematic areas. Panel~\ref{fig:mst}B reports the ratio between the two: it remains below one throughout and tends to decrease over time, confirming a progressive concentration. In the data, most of the expansion of the covered space occurs during the first few programmes, whereas the null model exhibits continued expansion over the whole period. Panels~\ref{fig:mst}C and \ref{fig:mst}D repeat the analysis using cumulative time windows aggregating all projects from FP1 up to each successive year. The cumulative comparison reinforces the yearly findings and reveals the trend more clearly: the ratio (panel~\ref{fig:mst}D) remains below zero and decreases steadily over time, showing that the gap between the observed and random MST length widens as the research portfolio matures, confirming the progressive thematic concentration.

We also perform this analysis separately on the embeddings of projects with at least one participant belonging to the \textit{Research} group and projects with at least one participant belonging to the \textit{Industry} group. These two groups only partially overlap, and we can therefore examine whether there are differences in the space coverage. We only use data from H2020 and Horizon Europe since they consistently report the type of organization, unlike the previous programmes. The results are reported in Appendix \ref{app:mst}. 
We do not appreciate a remarkable difference between the MST length of projects with academic or industrial partners, indicating that the presence of industry-related organizations does not strongly influence how the semantic space is covered.

The consistently sub-random MST length has important implications. Instead of exploring the semantic space uniformly as would be expected if new projects were positioned independently across all research directions, European research exhibits a focused exploration, with new projects clustering near existing thematic areas. This pattern is consistent with path-dependent dynamics in which established competencies, infrastructure, and collaboration networks channel new research into adjacent domains rather than into genuinely novel ones~\cite{mazzucato2018,arthur1989}. The fact that most semantic expansion occurs during the earlier FPs, with progressively less in later programmes, suggests a form of thematic maturation: the European research portfolio establishes its broad contours early and subsequently fills in and refines these areas rather than expanding into fundamentally new territory.

This finding carries direct policy relevance. If FPs are intended to support transformational, mission-oriented research~\cite{kuhlmann2018}, the observed pattern of focused concentration raises questions about whether the current funding architecture provides sufficient incentives for genuine exploration beyond established thematic boundaries. The contrast with the null model where expansion continues throughout the observation period illustrates the magnitude of this path dependence.

\subsection{Comparison academia and industry}
\label{sec:res_ind}

We now examine how funding is distributed between research organizations and industry across thematic domains. As described in Section~\ref{sec:data}, we restrict this analysis to H2020 and Horizon Europe, the two programmes for which complete funding information is available. For each country, we construct two vectors of length 16 (one entry per macro-topic), containing the total funding (based on the \textit{netEcContribution} field) allocated to \textit{Research} and \textit{Industry} organizations, respectively.
In Figure~\ref{fig:corr_res_ind}, we report the results for the eight most populous countries in Europe (Germany, Great Britain, France, Italy, Spain, Poland, Romania, Netherlands) for a better visualization. The total matrix can be found in Appendix \ref{app:corr_matrix}. 

To quantify the alignment between research and industry funding profiles, we compute two types of Pearson correlation. The ``within-country'' correlation compares the research and industry vectors of the same country. The ``between-countries'' correlation compares, for each pair of countries, their research vectors (lower triangle of Figure~\ref{fig:corr_res_ind}) and their industry vectors (upper triangle). Several patterns emerge. Cross-country correlations are substantially higher for research than for industry, as indicated by the darker blue shading in the lower triangle. This asymmetry reveals that European research organizations have converged toward a relatively common set of thematic priorities, consistent with the homogenizing effects of shared evaluation criteria and programme structures~\cite{doussineau2020}, whereas industrial participation may remain shaped by nationally specific industrial structures and comparative advantages. This divergence has implications for policies aimed at strengthening the translation of publicly funded research into commercial innovation across different national contexts~\cite{foray2015}.

The diagonal reports the within-country research-industry correlation. Spain displays the highest value, suggesting a closer alignment between its academic research base and its industrial fabric. Rather than indicating a more balanced innovation ecosystem, however, this may reflect a more limited industrial scope that naturally aligns with the country's academic priorities.

\begin{figure}[H]
\centering
\includegraphics[width=1\textwidth]{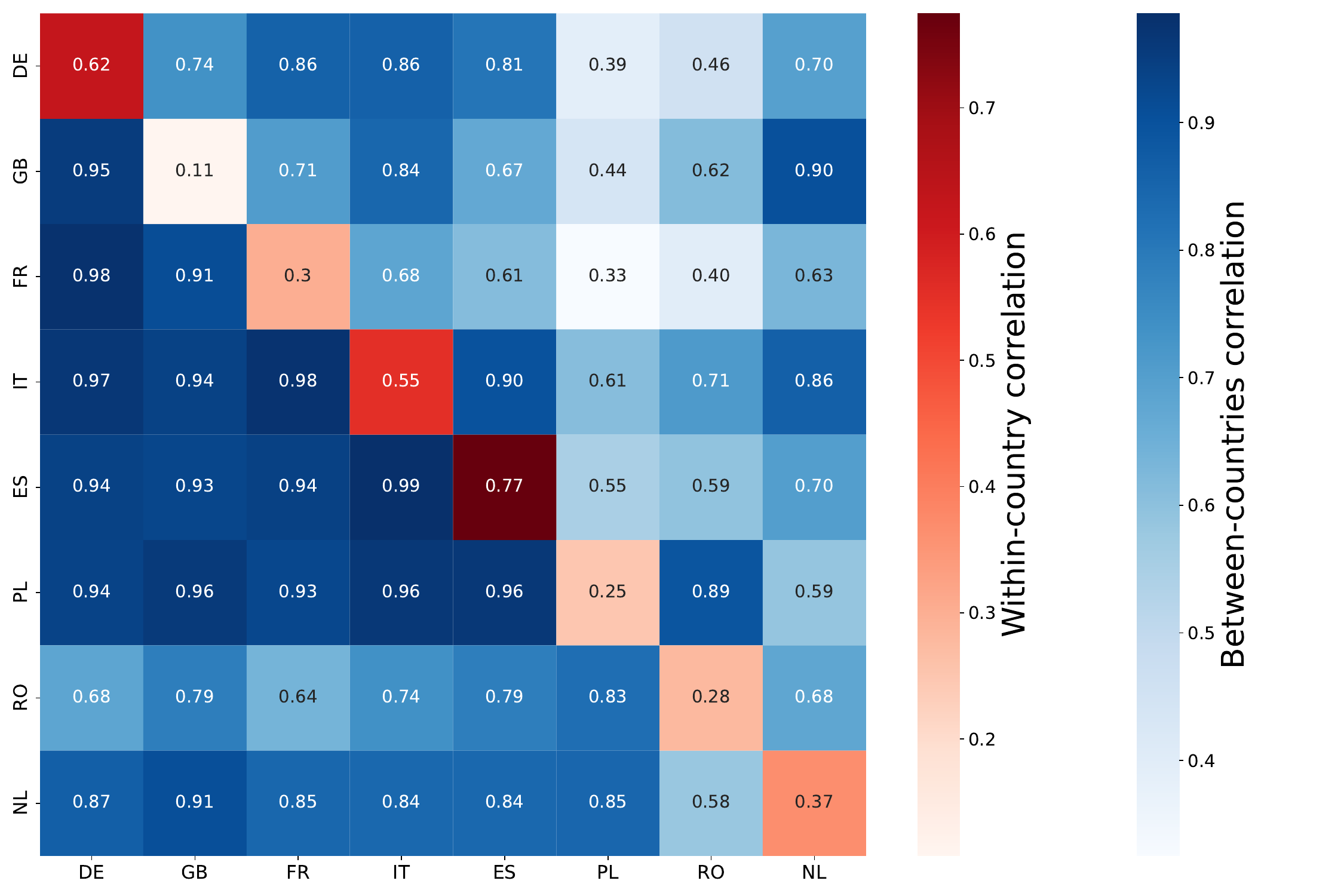}
\caption{\textbf{Correlation matrix of funding per topic per country (H2020 and Horizon Europe).} Correlation values for the 8 most populous countries in Europe are reported. The lower triangular part of the matrix shows the Pearson correlation coefficient, for each pair of countries, between their research funding vectors. The upper part, instead, between their industry funding vectors. Darker blue cells (more present in the lower part, corresponding to research) correspond to higher correlation. The diagonal, instead, reports the Pearson correlation coefficient, for each country, between their research and industry funding vectors. Darker red cells indicate higher correlation.}
\label{fig:corr_res_ind}
\end{figure}

\begin{figure}[H]
\centering
\includegraphics[width=1\textwidth]{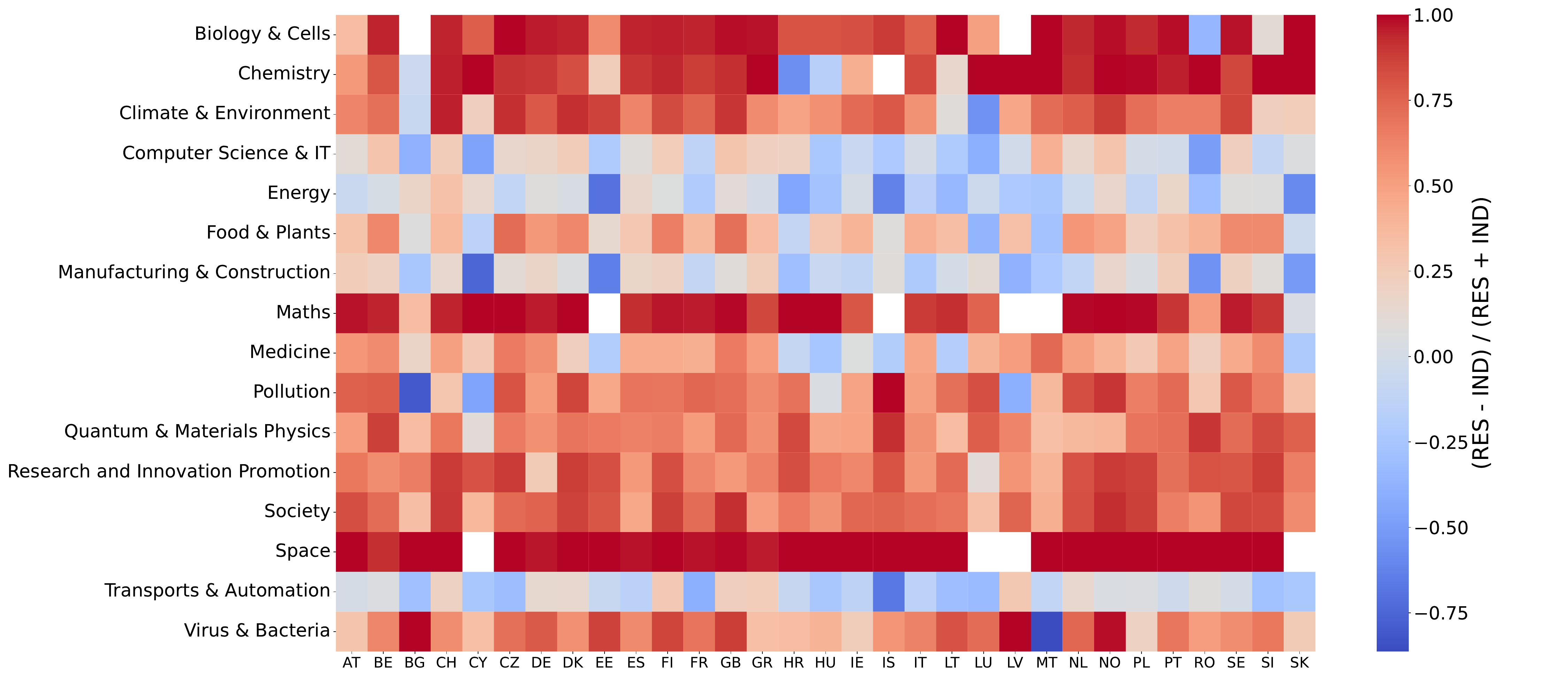}
\caption{\textbf{Money per macro-topic per country in research and industry (H2020 and Horizon Europe).} For each macro-topic, for each country, the value $(RES-IND)/(RES+IND)$ is reported, where $RES$ and $IND$ respectively indicate how much money per capita is assigned to organizations belonging to \textit{Research} and \textit{Industry}. Blank cells indicate that for a specific country there are no funded projects classified in the corresponding macro-topic.}
\label{fig:country_topics}
\end{figure}

Figure~\ref{fig:country_topics} provides a complementary, topic-level view. For each country and macro-topic, it reports the normalized balance $(RES-IND)/(RES+IND)$, where $RES$ and $IND$ denote the per-capita funding allocated to \textit{Research} and \textit{Industry} organizations, respectively. Red cells indicate research-dominated funding, blue cells industry-dominated funding. Across most topics and countries, funding flows predominantly to research. However, several applied domains, notably \textit{Energy}, \textit{Manufacturing \& Construction}, and \textit{Transports \& Automation}, display a more balanced or industry-leaning profile, reflecting the close intertwining of applied research and technological deployment in these fields. By contrast, domains such as \textit{Biology \& Cells}, \textit{Virus \& Bacteria}, and \textit{Maths} exhibit strongly research-dominated profiles, consistent with their more fundamental scientific orientation. 

\section{Discussion}
\label{sec:discussion}

This study provides the first integrated longitudinal analysis of the European research collaboration network and the thematic evolution of its research portfolio across all nine FPs. In this section, we revisit our three research questions in the light of the empirical findings and discuss their implications for EU science and innovation policy.

\subsection{RQ1: Integration without full convergence}

Our first research question asked whether the progressive expansion of the European collaboration network reduces structural inequalities and how integration evolves for incumbent and newly entering countries. The evidence points to a qualified affirmative: FPs do serve their intended function as instruments of integration, but the process is neither instantaneous nor complete.

The entry of new countries into the collaboration network temporarily increases inequality and reduces integration, but both effects are progressively mitigated as newcomers establish collaborative relationships. The within-cohort decomposition is particularly revealing: for each successive group of entrants, the Gini index drops markedly in the programmes following their accession, and integration rises steadily. This suggests that sustained participation in the FP ecosystem generates a genuine convergence dynamic. However, the slight increase in inequality observed in the most recent programmes, even as integration continues to improve, echoes concerns raised in the ``widening'' debate and indicates that the competitive dynamics of excellence-based funding may be generating new forms of stratification even as older disparities diminish.

The gravity model results add further nuance. The low importance of geographical distance as a predictor of collaboration, combined with the dominant role of population and GDP per capita, suggests that the barriers to European collaboration are more structural rather than geographical. Countries participate less because of where they are and more because of the size and strength of their research systems, a finding with direct implications for the design of instruments aimed at promoting cohesion. Widening policies that focus on removing geographical barriers may therefore be less effective than interventions that address the structural capacity of national research systems.

\subsection{RQ2: Path dependence in the research portfolio}

Our second research question asked whether the thematic landscape of European research exhibits genuine exploratory diversification or follows a path-dependent trajectory. The MST analysis provides a clear answer: European research explores its semantic space in a focused rather than uniform manner, and this concentration intensifies over time.

The MST length is consistently shorter than expected under the null model, and the ratio between the two decreases over successive programmes. Most of the expansion of the thematic space occurs during the earlier FPs, with progressively less thereafter. This pattern is consistent with theoretical accounts of path dependence in which established competencies, institutional structures, and network effects channel research into incremental extensions of existing programmes rather than radical departures~\cite{mazzucato2018,arthur1989}.

For a funding system that aspires to support mission-oriented, transformative research~\cite{kuhlmann2018}, this finding raises important questions. The observed thematic maturation could be interpreted positively as evidence that European research has consolidated deep expertise in key areas, or as a warning that the current programme architecture insufficiently rewards genuine exploration. The contrast between the observed pattern and the null model quantifies the magnitude of this path dependence and provides a baseline against which the effects of future policy interventions could be measured.

\subsection{RQ3: Research-industry alignment and national variation}

Our third research question concerned the balance between research and industry funding across thematic domains and countries. The analysis reveals a structural asymmetry in the European innovation ecosystem that operates along two distinct dimensions.

Along the cross-country dimension, European research organizations have converged toward relatively similar thematic priorities, a likely consequence of shared evaluation frameworks and programme structures~\cite{doussineau2020}, while industrial participation remains nationally differentiated. This divergence suggests that the ``translation gap'' between publicly funded research and commercial application varies substantially across national contexts, and that a pan-European innovation policy may need to be more sensitive to these differences than current programme designs allow~\cite{foray2015}.

Along the thematic dimension, the balance between research and industry funding follows an expected but informative pattern. Applied domains such as \textit{Energy}, \textit{Manufacturing \& Construction}, and \textit{Transports \& Automation} display balanced or industry-leaning funding profiles, reflecting the close intertwining of applied research and technological deployment in these fields. Fundamental domains such as \textit{Biology \& Cells}, \textit{Maths}, and \textit{Virus \& Bacteria} show strongly research-dominated profiles. The most informative aspect, however, is the country-level heterogeneity within each domain: the same thematic area can be predominantly research-funded in one country and industry-funded in another, suggesting that national innovation systems interact with European funding structures in distinctive ways, mediating the impact of FP investments on different economies.

\subsection{Limitations}

Several limitations should be acknowledged. First, the topic modeling results depend on specific methodological choices: embedding model, clustering parameters, and hierarchical merging thresholds. We have verified that the main findings, particularly regarding MST dynamics and broad research-industry patterns, are robust to reasonable parameter variations, but the exact delineation of macro-topics should be interpreted with appropriate caution.

In addition, our analysis of Horizon Europe is necessarily incomplete, as we could only include projects starting before January 2026, representing approximately half of the programme's planned duration. The most recent data point may therefore not fully reflect the thematic and structural dynamics of the current programme.

Moreover, the CORDIS database presents inconsistencies in data format across different FPs, particularly regarding organization identifiers and funding information. Our country-level analysis mitigates many of these issues, but finer-grained analyses at the organizational or regional level would require more extensive data harmonization.

Furthermore, the gravity model provides useful descriptive insights, but is also a simplification of the complex factors driving international research collaboration. We adopted a more general approach by using the geographical distance between countries centroids, but other more specific measures such as border distance \cite{head2014gravity} or flight distance \cite{catalini2020travel} can be explored. Moreover, cultural proximity, linguistic ties, historical relationships, and discipline-specific collaboration patterns are not captured by our three-variable specification.

Finally, the semantic space analysis operates on the full high-dimensional embedding space. The relationship between geometric proximity in this space and substantive intellectual proximity deserves further investigation, as the properties of high-dimensional embeddings do not always correspond intuitively to human assessments of thematic similarity.

\subsection{Implications and future research}

Taken together, our answers to the three research questions suggest that the FP architecture has been largely effective in its integrative mission (RQ1), but faces a tension between consolidation and exploration (RQ2) that is further complicated by the nationally differentiated nature of industrial engagement (RQ3). These findings carry several implications for the design of future FPs. The evidence of progressive integration lends support to the long-term strategic value of ``widening'' instruments. However, the path-dependent evolution of the research portfolio and the slight recent increase in inequality suggest that additional mechanisms may be needed to prevent the consolidation of a two-tier system and to promote genuinely exploratory research.

Future research could extend this analysis in several directions. A more granular analysis at the organizational or regional level could reveal heterogeneities within countries that are masked by our country-level approach. The integration of bibliometric data such as publications and citations arising from FP-funded projects would allow the assessment of research impact and knowledge flows associated with different network positions and thematic areas. The application of causal inference methods, for instance exploiting the quasi-experimental variation generated by funding thresholds or programme transitions, could help establish whether the patterns we observe are driven by the FP architecture itself or by broader trends in European research. Finally, extending the semantic analysis to track how specific thematic areas emerge, grow, and potentially decline would provide a more complete picture of the lifecycle of research priorities in the European context.

\section{Conclusion}
\label{sec:conclusion}

This study has provided the first integrated longitudinal analysis of the European research collaboration network and the thematic evolution of its research portfolio across all nine FPs, from FP1 in 1984 to the ongoing Horizon Europe. By combining structural network analysis with semantic characterization of over 120,000 project descriptions, we have shown that FPs produce progressive but incomplete integration (RQ1), path-dependent thematic concentration rather than broad exploratory diversification (RQ2), and a convergence in public research priorities that coexists with nationally differentiated industrial engagement (RQ3).

These three findings are distinct but connected by a common thread. The FP system has proven remarkably effective at building a pan-European collaborative infrastructure: geographical distance has become irrelevant as a predictor of collaboration, newcomers are progressively absorbed into the network, and research organizations across the continent have converged toward shared thematic priorities. Yet this very success carries a structural risk. The same mechanisms that foster integration (shared evaluation criteria, cumulative network effects, the gravitational pull of established research hubs) also channel activity along well-trodden paths, making it harder for the system to venture into genuinely unexplored territory. The tension is  between integration and exploration more than between integration and fragmentation: the more cohesive the European research system becomes, the more it may tend toward thematic consolidation.

Such tension is not unique to the FPs. It echoes a broader dilemma in science and innovation policy between exploitation of existing strengths and exploration of new possibilities. What our analysis contributes is an empirical measure of this trade-off at a continental scale and over a four-decade horizon. The MST analysis, in particular, provides a quantitative baseline (the gap between observed and random thematic expansion) that could serve as a diagnostic tool for the monitoring of whether future policy interventions succeed in shifting the balance toward greater exploration.

Looking ahead, the design of the successor to Horizon Europe offers an opportunity to address the dynamics documented here. Three considerations emerge from our analysis. First, the evidence that integration is a slow, cumulative process, requiring multiple programme cycles to take effect, argues for long-term commitment to widening instruments rather than short-term corrective interventions. Second, the path-dependent nature of thematic evolution suggests that promoting transformative research may require dedicated funding mechanisms that are structurally insulated from the gravitational pull of established priorities, and not merely additional calls within the existing architecture. Third, the divergence between convergent research priorities and nationally differentiated industrial participation implies that a one-size-fits-all approach to research-to-innovation translation is unlikely to be effective; programme design may need to accommodate, rather than override, the diversity of national innovation systems.

More broadly, the forty-year trajectory of the FPs offers a case study in how large-scale policy instruments evolve. What began as a relatively modest experiment in transnational R\&D cooperation has grown into a defining feature of the European research landscape, shaping not only who collaborates with whom, but also what is investigated and how public and private research agendas relate to each other. The longitudinal perspective adopted in this study reveals that the effects of such instruments are cumulative, path-dependent, and often slow to materialize. These features are easily missed by evaluations focused on single programme cycles. As the European Union enters a period of reflection on the future of its research policy, an empirically grounded understanding of these long-run dynamics is essential for designing funding architectures that are capable of fostering both cohesion and the capacity for renewal.


\section*{Data availability}
The code to reproduce the results in the paper can be found in the repository \url{https://github.com/vorsanigo/EU-project}, while the useful datasets at the following link \url{https://osf.io/9u2m8/files/osfstorage}.

\subsection*{Acknowledgements}
The authors thank Alberto Amaduzzi and Ferruccio Resta for the insightful discussions.

\clearpage
\appendix

\section{Appendix: Table of 32 countries}
\label{app:table_countries}

\begin{table}[h]
\begin{center}
\begin{tabular}{ p{2.5 cm} p{3 cm}}
\hline
\textbf{Country} & \textbf{ISO 2 code} \\
\hline
Austria & AT \\
Belgium & BE \\
Bulgaria & BG \\
Cyprus & CY \\
Croatia & HR \\
Czechia & CZ \\
Denmark & DK \\
Estonia & EE \\
Finland & FI \\
France & FR \\
Germany & DE \\
Hungary & HU \\
Iceland & IS \\
Ireland & IE \\
Italy & IT \\
Latvia & LV \\
Liechtenstein & LI \\
Lithuania & LT \\
Luxembourg & LU \\
Malta & MT \\
Netherlands & NL \\
Norway & NO \\
Poland & PL \\
Portugal & PT \\
Romania & RO \\
Slovakia & SK \\
Slovenia & SI \\
Spain & ES \\
Sweden & SE \\
Switzerland & CH \\
Great Britain & GB \\
Greece & GR \\
\hline
\end{tabular}
\end{center}
\caption{\textbf{Table of the 32 countries considered for the study.} The table reports the list of the 32 countries (EU + EFTA + UK) considered in this study with the corresponding ISO 2 code.}
\label{tab:data}
\end{table}

\clearpage
\begingroup
\small
\section{Appendix: Organizations' location inference}
\label{app:geo_city}
The \textit{Organizations} dataset does not report the city (field \textit{city}) or, more generally, the geographical information (eg: latitude and longitude) of all the organizations. For this reason, we implemented a pipeline to determine the city and consequently latitude and longitude of the organizations for which the information was missing. We based it on the field \textit{city} and we adopted different strategies and tools, which include the geocoding system \textit{Nominatim}\footnote{\url{https://nominatim.org/}}, the computation of \textit{Levenshtein distance}\footnote{Levenshtein, V. I. (1966, February). Binary codes capable of correcting deletions, insertions, and reversals. In Soviet physics doklady (Vol. 10, No. 8, pp. 707-710).} between strings ($d$), web scraping, and the model \texttt{gpt-4o-mini}\footnote{\url{https://platform.openai.com/docs/models}} from OpenAI.

Let us call $O$ the set of all the organizations, $O_m$ the set of the organizations with no associated city, $O_{nf}$ the set of organizations with a city string associated which has no correspondence when sent to Nominatim, $C_{nf}$ the set of such city strings, $O_{ok}$ the set of organizations that have a city associated found also by Nominatim, and $C_{ok}$ the set of these cities as returned by Nominatim. The idea is to find the cities of the organizations and to homogenize their names according to how they are defined in Nominatim. The pipeline to retrieve the cities of the organizations consists of a series of steps presented below.

Note that when sending a city string to Nominatim, it returns both the city and its country, so we compare the country with the one where the organization is located according to the \textit{Organizations} dataset to have a ground truth for our results, which we also manually check when they are uncertain.

\begin{algorithm}[H]
\caption{Organizations' City Retrieval Pipeline}

\KwIn{Set of organizations $O$}
\KwOut{Organizations with validated cities $O_{ok}$}

\For{$o \in O$}{

    \If{$o$ has city $c \in C_{ok}$}{
        Add $o$ to $O_{ok}$
    }

    \ElseIf{$o$ has no city associated}{
        Add $o$ to $O_m$

        Query \texttt{gpt-4o-mini} for city $c$ of $o$

        Send $c$ to Nominatim

        \If{Nominatim returns $c_n \neq \text{None}$}{
            Assign $c_n$ to $o$

            Add $o$ to $O_{ok}$

            Add $c_n$ to $C_{ok}$
        }
    }

    \ElseIf{$o$ has city associated $c \in C_{nf}$}{

        Add $o$ to $O_{nf}$

        Compute $d(c,c_{ok})$ for all $c_{ok} \in C_{ok}$

        Let $c^* = \arg\min_{c_{ok} \in C_{ok}} d(c,c_{ok})$

        \If{$d(c,c^*) \leq t$ (with $t=3$)}{

            Assign $c^*$ to $o$

            Add $o$ to $O_{ok}$

        }

        \Else{

            $c_{web} \leftarrow \text{None}$

            Extract candidate city $c_{google}$ from Google

            \If{$c_{google} \neq \text{None}$}{
                $c_{web} \leftarrow c_{google}$
            }

            \Else{

                Extract candidate city $c_{wiki}$ from Wikipedia

                \If{$c_{wiki} \neq \text{None}$}{
                    $c_{web} \leftarrow c_{wiki}$
                }

            }

            \If{$c_{web} \neq \text{None}$}{

                Send $c_{web}$ to Nominatim

                \If{Nominatim returns $c_n \neq \text{None}$}{

                    Assign $c_n$ to $o$

                    Add $o$ to $O_{ok}$

                    Add $c_n$ to $C_{ok}$
                }

            }

            \Else{

                Query \texttt{gpt-4o-mini} for city $c$ of $o$

                Send $c$ to Nominatim

                \If{Nominatim returns $c_n \neq \text{None}$}{

                    Assign $c_n$ to $o$

                    Add $o$ to $O_{ok}$

                    Add $c_n$ to $C_{ok}$
                }

            }

        }

    }

}

\end{algorithm}

\clearpage
\section{Appendix: Minimum Spanning Tree: research-related VS industry-related organizations}
\label{app:mst}
\begin{figure}[h]
\centering
\includegraphics[width=0.9\textwidth]{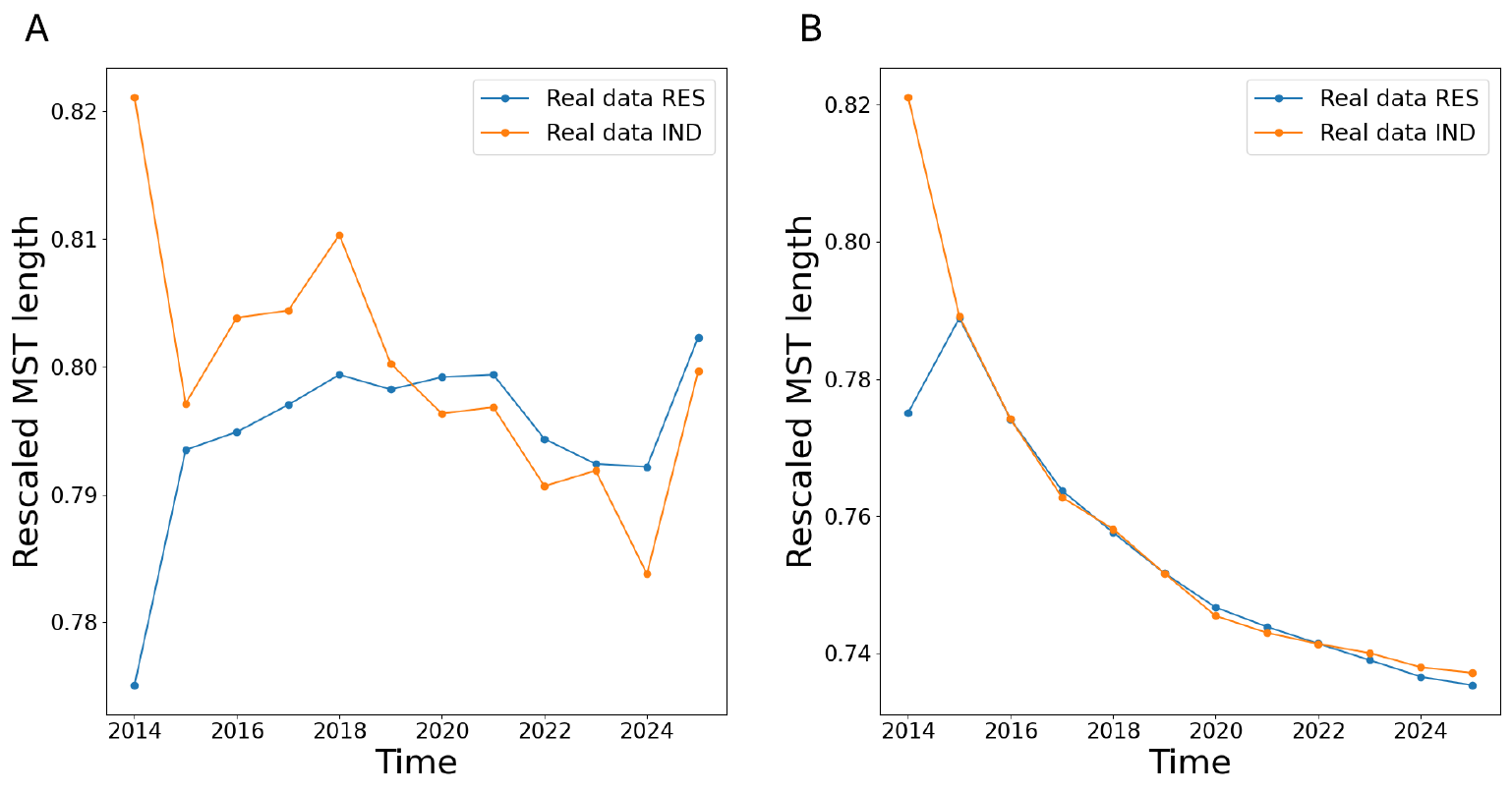}
\caption{\textbf{Temporal Minimum Spanning Tree.} We report the computed rescaled MST length in different cases. The measure is obtained by dividing the MST length by $n^{(d-1)/d}$, where $n$ is the number of points and $d=1024$ is the embedding dimension. \textbf{(A)} MST length computed on time windows of 1 year on different subsets of project embeddings: projects with at least one participant belonging to the group \textit{Research} (in blue) VS projects with at least one participant belonging to the group \textit{Industry} (in orange). \textbf{(B)} Cumulative MST length computed on different subsets of project embeddings: projects with at least one participant belonging to the group \textit{Research} (in blue) VS projects with at least one participant belonging to the group \textit{Industry} (in orange).}

\label{fig:corr_res_ind}
\end{figure}

\clearpage
\endgroup
\section{Appendix: Correlation matrix full}
\label{app:corr_matrix}
\begin{figure}[h]
\centering
\includegraphics[width=0.79\textwidth]{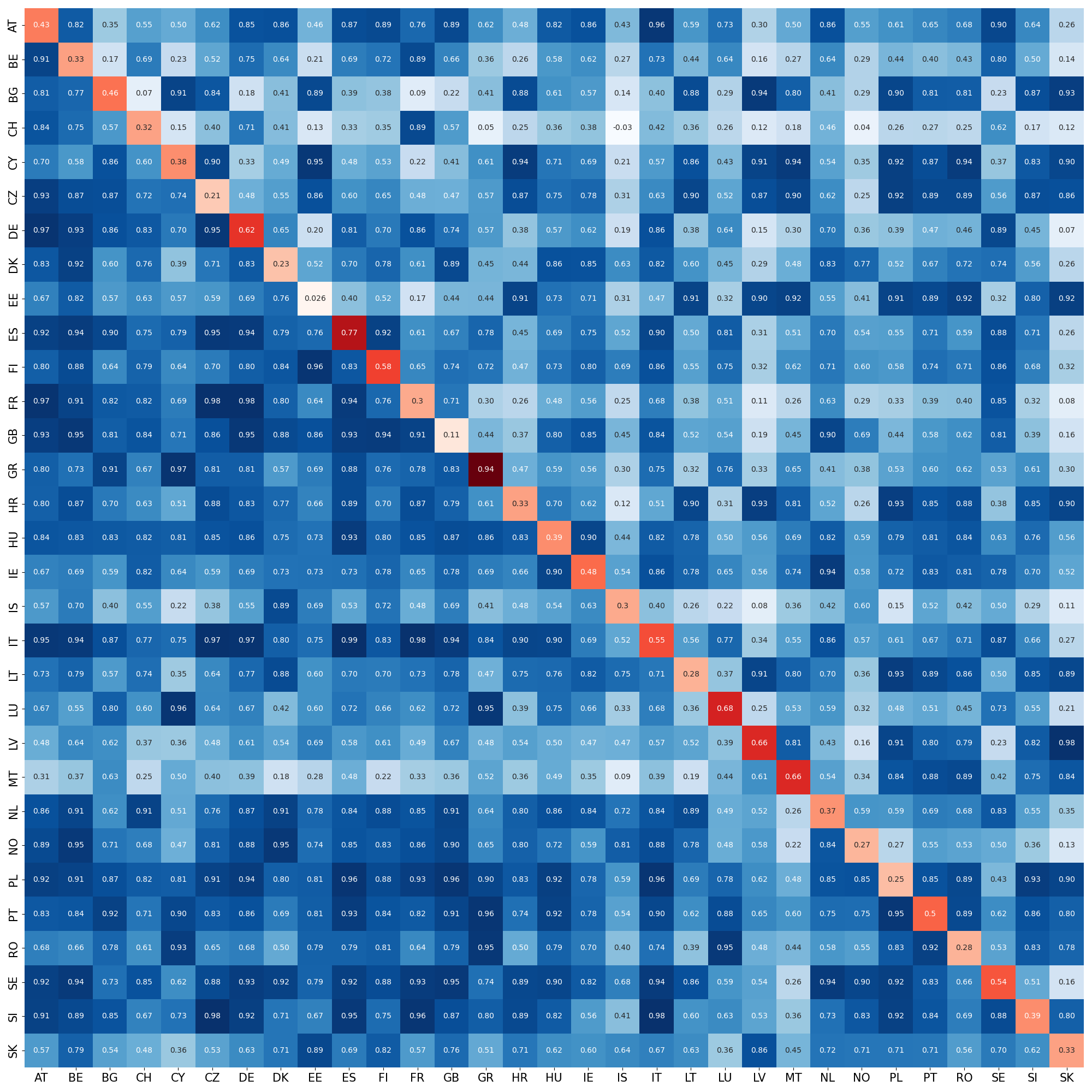}
\caption{\textbf{Correlation matrix of funding per topic per country.} The lower triangular part of the matrix shows the Pearson correlation coefficient, for each pair of countries, between their research funding vectors. The upper part, instead, between their industry funding vectors. Darker blue cells (more present in the lower part, corresponding to research) correspond to higher correlation. The diagonal, instead, reports the Pearson correlation coefficient, for each country, between their research and industry funding vectors. Darker red cells indicate higher correlation.}
\label{fig:corr_res_ind}
\end{figure}

\clearpage
\bibliographystyle{unsrtnat}
\bibliography{references}
\end{document}